\documentclass[12pt]{article}
\usepackage{amsfonts}

\usepackage{amsmath}

\newtheorem{theorem}{Theorem}
\newtheorem{acknowledgement}[theorem]{Acknowledgement}

\newtheorem{corollary}[theorem]{Corollary}

\newtheorem{proposition}[theorem]{Proposition}
\newtheorem{remark}[theorem]{Remark}

\newenvironment{proof}[1][Proof]{\textbf{#1.} }{\ \rule{0.5em}{0.5em}}
\input{tcilatex}

\begin{document}

\title{Quaternionic fundamental solutions for the numerical analysis of
electromagnetic scattering problems}
\author{Kira V. Khmelnytskaya, Vladislav V. Kravchenko \and and \and %
Vladimir S. Rabinovich \\
Depto. de Telecomunicaciones, SEPI\\
Escuela Superior de Ingenier\'{\i}a Mec\'{a}nica y El\'{e}ctrica\\
Instituto Polit\'{e}cnico Nacional\\
C.P.07738 M\'{e}xico D.F., \\
MEXICO}
\maketitle

\begin{abstract}
We propose a new class of fundamental solutions for the numerical analysis
of boundary value problems for the Maxwell equations. We prove completeness
of systems of such fundamental solutions in appropriate Sobolev spaces on a
smooth boundary and support the relevancy of our approach by numerical
results.


\end{abstract}

\section{Introduction}

The method of fundamental solutions or which is the same of discrete sources
(we will keep to the first name) is a widely used technique for the
numerical solution of elliptic boundary value problems which falls in the
class of so called boundary methods reducing problems in $n$-dimensional
domains to some equations on their $(n-1)$-dimensional boundaries. It is
applicable when a fundamental solution of the differential equation of the
problem is known and the completeness of an infinite system of such
fundamental solutions with singularities (sources) placed outside the domain
of the problem is proved. The original idea of the method emerged in sixties %
\cite{Browder, Kupr, KupAl} and since then the method of fundamental
solutions was successfully used in geophysics, acoustics, elasticity theory,
electromagnetism and other fields. We refer the reader to the books \cite%
{Alexbook, Eremin} and to the review \cite{Karag} for bibliography and more
information about the method.

The aim of the present paper is to propose necessary elements for the
application of the method of fundamental solutions to boundary value
problems of electromagnetic scattering theory, in particular, to introduce a
system of fundamental solutions \ for Maxwell's equations and to prove its
completeness in appropriate functional spaces. Here an explanation is needed
because all this seems to represent nothing new. Let us start with the
concept of a fundamental solution for a system of partial differential
equations with $n$ unknowns and $n$ equations of the form $Au=0$, where $A$
is a differential operator. Usually (see, e.g., \cite[p. 179]{Nedelec}) it
is defined as a matrix $n\times n$, denoted by $\Phi $, such that%
\begin{equation}
A\Phi =\delta E_{n},  \label{fundmatr}
\end{equation}%
where $\delta $ is the Dirac delta function and $E_{n}$ is the $n\times n$
identity matrix. Nevertheless such a definition has no clear physical
interpretation as a field generated by a point source, the usual meaning of
the fundamental solution. In this sense the electromagnetic fields produced
by an electric and a magnetic dipoles are closer to the physical meaning of
a fundamental solution and sometimes they are called the fundamental
solutions of the Maxwell system \cite[Sect. 4.2]{CK1}, but then it is not
clear how can they be used for the analytical solution of homogeneous and
inhomogeneous Maxwell's equations, usually based on property (\ref{fundmatr}%
).

We propose another possibility, a fundamental solution which enjoys both
properties. It satisfies (\ref{fundmatr}) in a sense explained below and has
a clear meaning of a field generated by a point source. Moreover, we prove
the completeness of an infinite system of such fundamental solutions in
appropriate Sobolev spaces which makes it possible to apply our system to
the numerical solution of boundary value problems for Maxwell's equations in
chiral media.

The construction of the system of fundamental solutions for the Maxwell
equations proposed here is based on some elements of quaternionic analysis
which seems to be the most appropriate formalism for this task. The
solutions obtained are complex quaternions, that is instead to be a pair of
three-component vectors, they have four components. Due to their lower
singularity and simple form the numerical application of them is easier and
more natural compared with solutions based on a matrix approach (see \cite%
{Alexbook, Eremin}).

The main idea to obtain the quaternionic fundamental solutions for Maxwell's
equations consists in the quaternionic diagonalization of Maxwell's
equations proposed in \cite{Krdep} (see also \cite{KSbook, Krdiag}). The
Maxwell equations for an isotropic homogeneous medium are reduced to a pair
of quaternionic equations in which the unknown functions are separated. For
each of these equations a fundamental solution is easily constructed and
then linear combinations of them will give the required system of
fundamental solutions for Maxwell's equations. The main difficulty
constitutes the proof of completeness of this system. We base our proof on
the completeness of a system of fundamental solutions of the Helmholtz
operator in the kernel of this operator in $L_{2}$-norm and make use of a
quaternionic decomposition of the kernel of the Helmholtz operator.

Our results are applied to Maxwell's equations for chiral media but they are
completely new for a nonchiral case as well.

\section{Complex quaternions}

We shall denote by $\mathbb{H}(\mathbb{C})$ the set of complex quaternions
(= biquaternions) (the letter $\mathbb{H}$ is frequently chosen in honor of
the inventor of quaternions, W. R. Hamilton). Each element $a$ of $\mathbb{H}%
(\mathbb{C})$ is represented in the form $a=\sum_{k=0}^{3}a_{k}i_{k}$ where $%
\{a_{k}\}\subset \mathbb{C}$, $i_{0}$ is the unit and $\{i_{k}|\quad
k=1,2,3\}$ are the quaternionic imaginary units, that is the standard basis
elements possessing the following properties:

\begin{equation*}
i_{0}^{2}=i_{0}=-i_{k}^{2};\;i_{0}i_{k}=i_{k}i_{0}=i_{k},\quad k=1,2,3;
\end{equation*}

\begin{equation*}
i_{1}i_{2}=-i_{2}i_{1}=i_{3};\;i_{2}i_{3}=-i_{3}i_{2}=i_{1};%
\;i_{3}i_{1}=-i_{1}i_{3}=i_{2}.
\end{equation*}%
We denote the imaginary unit in $\mathbb{C}$ by $i$ as usual. By definition\
\ $i$\ commutes with \ $i_{k}$, $k=\overline{0,3}$.

The basic quaternionic imaginary units $i_{1}$, $i_{2}$ and $i_{3}$ can be
identified with the basic coordinate vectors in a three-dimensional space.
In this way a vector $\overrightarrow{a}$ from $\mathbb{C}^{3}$ is
identified with the complex quaternion $a_{1}i_{1}+a_{2}i_{2}+a_{3}i_{3}$.
We will use the so called vector representation of complex quaternions, each 
$a\in \mathbb{H}(\mathbb{C})$ is represented as follows $a=a_{0}+%
\overrightarrow{a}$, where $a_{0}$ is the scalar part of $a$ sometimes
denoted as $\limfunc{Sc}(a)=a_{0}$ and $\overrightarrow{a}$ is the vector
part of $a$: $\limfunc{Vec}(a)=\overrightarrow{a}$ $%
=\sum_{k=1}^{3}a_{k}i_{k} $. Complex quaternions of the form $a=%
\overrightarrow{a}$ will be called purely vectorial.

In vector terms, the multiplication of\ two\ arbitrary\ complex\
quaternions\ $a$ and $b$\ can\ be\ rewritten\ as\ follows:\ 

\begin{equation*}
a\cdot b=a_{0}b_{0}-<\overrightarrow{a},\overrightarrow{b}>+\left[ 
\overrightarrow{a}\times \overrightarrow{b}\right] +a_{0}\overrightarrow{b}%
+b_{0}\overrightarrow{a},
\end{equation*}%
where\ 

\begin{equation*}
<\overrightarrow{a},\overrightarrow{b}>:={\displaystyle\sum_{k=1}^{3}}%
a_{k}b_{k}\in\mathbb{C},
\end{equation*}

\begin{equation*}
\lbrack \overrightarrow{a}\times \overrightarrow{b}]:=\left| 
\begin{array}{lll}
i_{1} & i_{2} & i_{3} \\ 
a_{1} & a_{2} & a_{3} \\ 
b_{1} & b_{2} & b_{3}%
\end{array}%
\right| \in \mathbb{C}^{3}.
\end{equation*}

The complex quaternion $\overline{a}=a_{0}-\overrightarrow{a}$ is called the
conjugate of $a$. Let us denote by $\mathfrak{S}$ the set of zero divisors
from $\mathbb{H}(\mathbb{C})$. We recall (see, e.g., \cite{KSbook}) that $%
a\in \mathfrak{S}$ iff $a\cdot \overline{a}=0$. If $a\notin \mathfrak{S}\cup
\left\{ 0\right\} $ then $a^{-1}$ exists and $a^{-1}=\overline{a}/(a\cdot 
\overline{a})$.

\section{Quaternionic differential operators}

We shall consider $\mathbb{H}(\mathbb{C})$-valued functions depending on
three variables $x_{1}$, $x_{2}$ and $x_{3}$. On the set of such
componentwise continuously differentiable functions the operator $%
D=\sum_{k=1}^{3}i_{k}\partial _{k},$ where $\partial _{k}=\frac{\partial }{%
\partial x_{k}}$, is defined. The expression $Df$, where $\ f$ is an $%
\mathbb{H}(\mathbb{C})$-valued function, can be rewritten in a vector form
as follows

\begin{equation*}
Df=-\limfunc{div}\overrightarrow{f}+\limfunc{grad}f_{0}+\limfunc{rot}%
\overrightarrow{f}.
\end{equation*}%
That is, $\limfunc{Sc}(Df)=-\limfunc{div}\overrightarrow{f}$ and $\limfunc{%
Vec}(Df)=\limfunc{grad}f_{0}+\limfunc{rot}\overrightarrow{f}$. The condition 
$f\in \ker D$ is equivalent to the Moisil-Theodoresco system 
\begin{equation}
\left\{ 
\begin{array}{c}
\limfunc{div}\overrightarrow{f}=0,\quad \\ 
\limfunc{grad}f_{0}+\limfunc{rot}\overrightarrow{f}=0,%
\end{array}%
\right.  \label{MTsyst}
\end{equation}%
which has been studied in hundreds of works (see, e.g., \cite{Bitsadze}, %
\cite{Dzhuraev}).

Denote $D_{\alpha }=D+\alpha I$, where $\alpha $ is a complex constant and $%
I $ is the identity operator. As we will see in the subsequent pages, $%
\alpha $ has the meaning of a wave number. Having this in mind we will
assume that $\alpha \neq 0$ and $\limfunc{Im}\alpha \geq 0$.

We have the following factorization of the Helmholtz operator%
\begin{equation}
\Delta +\alpha ^{2}=-D_{\alpha }D_{-\alpha }=-D_{-\alpha }D_{\alpha },
\label{factor}
\end{equation}%
which in particular means that any function satisfying the equation 
\begin{equation}
D_{\alpha }f=0  \label{D+}
\end{equation}%
or 
\begin{equation}
D_{-\alpha }f=0  \label{D-}
\end{equation}%
also satisfies the Helmholtz equation 
\begin{equation}
(\Delta +\alpha ^{2})f=0.  \label{Helm}
\end{equation}%
We will use the fundamental solution of the Helmholtz operator%
\begin{equation*}
\theta _{\alpha }(x)=-\frac{e^{i\alpha \left| x\right| }}{4\pi \left|
x\right| }
\end{equation*}%
which fulfills the Sommerfeld radiation condition at infinity.

Fundamental solutions $\mathcal{K}_{\alpha }$ and $\mathcal{K}_{-\alpha }$
for the operators $D_{\alpha }$ and $D_{-\alpha }$ respectively can be
obtained easily using (\ref{factor}). We have that the functions%
\begin{equation*}
\mathcal{K}_{\alpha }=-(D-\alpha )\theta _{\alpha }\qquad \text{and\qquad }%
\mathcal{K}_{-\alpha }=-(D+\alpha )\theta _{\alpha }
\end{equation*}%
satisfy the equations%
\begin{equation*}
D_{\pm \alpha }\mathcal{K}_{\pm \alpha }=\delta .
\end{equation*}%
More explicitly we have%
\begin{equation*}
\mathcal{K}_{\pm \alpha }(x)=(\pm \alpha +\frac{x}{\left| x\right| ^{2}}%
-i\alpha \frac{x}{\left| x\right| })\theta _{\alpha }(x),
\end{equation*}%
where $x=\sum_{k=1}^{3}x_{k}i_{k}$. Note that $\mathcal{K}_{\pm \alpha }(x)$
are complex quaternionic functions with $\limfunc{Sc}(\mathcal{K}_{\pm
\alpha }(x))=\pm \alpha \theta _{\alpha }(x)$ and $\limfunc{Vec}(\mathcal{K}%
_{\pm \alpha }(x))=-\limfunc{grad}\theta _{\alpha }(x)=(\frac{x}{\left|
x\right| ^{2}}-i\alpha \frac{x}{\left| x\right| })\theta _{\alpha }(x)$.

Let us introduce the following operators%
\begin{equation*}
\Pi _{\pm \alpha }=\mp \frac{1}{2\alpha }D_{\mp \alpha }
\end{equation*}%
considering them on $\mathbb{H}(\mathbb{C})$-valued functions from $\ker
(\Delta +\alpha ^{2})$. Then we have the following statement (see the proof
in \cite[p. 36]{KSbook}).

\begin{proposition}
\label{kerHelm} The following relations hold:

\begin{enumerate}
\item $\Pi _{\pm \alpha }^{2}=\Pi _{\pm \alpha };$

\item $\Pi _{\alpha }\Pi _{-\alpha }=\Pi _{-\alpha }\Pi _{\alpha }=0;$

\item $\Pi _{\alpha }+\Pi _{-\alpha }=I;$

\item as $\func{im}\Pi _{\pm \alpha }=\ker D_{\pm \alpha }$ we have 
\begin{equation*}
\ker (\Delta +\alpha ^{2})=\ker D_{\alpha }\oplus \ker D_{-\alpha }.
\end{equation*}
\end{enumerate}
\end{proposition}

\section{Quaternionic integral operators}

Let $\Gamma $ be a closed Liapunov surface in $\mathbb{R}^{3}$. The
corresponding interior domain we denote by $\Omega ^{+}$ and the exterior by 
$\Omega ^{-}$. Let $\overrightarrow{n}$ be the outward with respect to $%
\Omega ^{+}$ unitary normal on $\Gamma $ in quaternionic form: $%
\overrightarrow{n}=\sum\nolimits_{k=1}^{3}n_{k}i_{k}$. Denote%
\begin{equation*}
K_{\pm \alpha }f(x)=-\int_{\Gamma }\mathcal{K}_{\pm \alpha }(x-y)%
\overrightarrow{n}(y)f(y)d\Gamma _{y},\quad x\in \mathbb{R}^{3}\setminus
\Gamma ,
\end{equation*}%
where $f$ is an $\mathbb{H}(\mathbb{C})$-valued function and all the
products under the integral are quaternionic. The following important result
is well known (see, e.g., \cite[p. 70]{KSbook}).

\begin{theorem}
Let $f\in C^{1}(\Omega ^{+})\cap C(\overline{\Omega ^{+}})$ and $f\in \ker
D_{\pm \alpha }(\Omega ^{+})$. Then 
\begin{equation*}
f(x)=K_{\pm \alpha }f(x),\quad \forall x\in \Omega ^{+}.
\end{equation*}
\end{theorem}

\begin{remark}
In this paper the belonging of a complex quaternionic function $f$ to some
functional space means that each of its components $f_{k}$ belongs to that
space.
\end{remark}

For the consideration of equations (\ref{D+}) and (\ref{D-}) in the domain $%
\Omega ^{-}$ one needs appropriate radiation conditions at infinity. Such
conditions were introduced in \cite{MM} (see also \cite{KC2}). Solutions of (%
\ref{D+}) are required to satisfy the following equality uniformly in all
directions%
\begin{equation}
(1+\frac{ix}{\left| x\right| })\cdot f(x)=o(\frac{1}{\left| x\right| }%
),\qquad \text{when }\left| x\right| \rightarrow \infty .  \label{rad+}
\end{equation}%
For solutions of (\ref{D-}) in $\Omega ^{-}$ the corresponding radiation
condition has the form 
\begin{equation}
(1-\frac{ix}{\left| x\right| })\cdot f(x)=o(\frac{1}{\left| x\right| }%
),\qquad \text{when }\left| x\right| \rightarrow \infty .  \label{rad-}
\end{equation}%
Then we have the following result \cite{KC2}:

\begin{theorem}
\label{Cauchyext}Let $f\in C^{1}(\Omega ^{-})\cap C(\overline{\Omega ^{-}})$%
, $f\in \ker D_{\alpha }(\Omega ^{-})$ and satisfy the radiation condition (%
\ref{rad+}) or $f\in \ker D_{-\alpha }(\Omega ^{-})$ and satisfy (\ref{rad-}%
). Then 
\begin{equation*}
f(x)=-K_{\alpha }f(x)\qquad \text{or\qquad }f(x)=-K_{-\alpha }f(x)\quad 
\text{respectively, for any }x\in \Omega ^{-}.
\end{equation*}
\end{theorem}

As we will consider $\mathbb{H}(\mathbb{C})$-valued functions satisfying the
Helmholtz equation%
\begin{equation}
(\Delta +\alpha ^{2})u=0  \label{Helmeq}
\end{equation}%
in unbounded domains it will be convenient to obtain for such functions a
radiation condition at infinity in a quaternionic form. If a solution $u_{0}$
of the Helmholtz equation is a scalar function, then the corresponding
radiation condition is the well known Sommerfeld condition%
\begin{equation}
i\alpha u_{0}(x)-\left\langle \frac{x}{\left| x\right| },\limfunc{grad}%
u_{0}(x)\right\rangle =o(\frac{1}{\left| x\right| }),\qquad \text{when }%
\left| x\right| \rightarrow \infty .  \label{Som1}
\end{equation}%
For a vector solution $\overrightarrow{u}$ of the Helmholtz equation the
corresponding radiation condition has the form \cite[Sect. 4.2]{CK1}%
\begin{equation}
\left[ \limfunc{rot}\overrightarrow{u}\times \frac{x}{\left| x\right| }%
\right] +\frac{x}{\left| x\right| }\limfunc{div}\overrightarrow{u}-i\alpha 
\overrightarrow{u}=o(\frac{1}{\left| x\right| }),\qquad \text{when }\left|
x\right| \rightarrow \infty .  \label{Som2}
\end{equation}%
Let us notice that a vector solution $\overrightarrow{u}$ of the Helmholtz
equation fulfills this condition if and only if each Cartesian component of
\ $\overrightarrow{u}$ \ fulfills the Sommerfeld radiation condition %
\cite[Sect. 4.2]{CK1}. Thus, our quaternionic radiation condition must
include both (\ref{Som1}) for $u_{0}$ and (\ref{Som2}) for $\overrightarrow{u%
}$.

It is easy to obtain such a condition using Proposition \ref{kerHelm} and
radiation conditions (\ref{rad+}) and (\ref{rad-}). From Proposition \ref%
{kerHelm} we have that an $\mathbb{H}(\mathbb{C})$-valued solution $u=u_{0}+%
\overrightarrow{u}$ of (\ref{Helmeq}) has the form 
\begin{equation*}
u=\Pi _{\alpha }u+\Pi _{-\alpha }u,
\end{equation*}%
$\Pi _{\alpha }u$ fulfills (\ref{rad+}) and $\Pi _{-\alpha }u$ fulfills (\ref%
{rad-}). Thus we obtain%
\begin{equation*}
u(x)=-\frac{ix}{\left| x\right| }\cdot \Pi _{\alpha }u(x)+\frac{ix}{\left|
x\right| }\cdot \Pi _{-\alpha }u(x)+o(\frac{1}{\left| x\right| }),\qquad 
\text{when }\left| x\right| \rightarrow \infty .
\end{equation*}%
From the definition of $\Pi _{\pm \alpha }$ we have 
\begin{equation*}
2\alpha u(x)=\frac{ix}{\left| x\right| }\cdot (D-\alpha )u(x)+\frac{ix}{%
\left| x\right| }\cdot (D+\alpha )u(x)+o(\frac{1}{\left| x\right| }),\qquad 
\text{when }\left| x\right| \rightarrow \infty .
\end{equation*}%
Finally we arrive at the following radiation condition at infinity for
complex quaternionic solutions of the Helmholtz equation (\ref{Helmeq}):%
\begin{equation}
i\alpha u(x)+\frac{x}{\left| x\right| }\cdot Du(x)=o(\frac{1}{\left|
x\right| }),\qquad \text{when }\left| x\right| \rightarrow \infty .
\label{radHelm}
\end{equation}%
As is easy to see when $u=u_{0}$ the scalar part of this equality gives us
exactly the Sommerfeld condition (\ref{Som1}) and the vector part 
\begin{equation}
\left[ \frac{x}{\left| x\right| }\times \limfunc{grad}u_{0}(x)\right] =o(%
\frac{1}{\left| x\right| })  \label{redund1}
\end{equation}%
is a redundant equality because it is a simple consequence of the fact that
a scalar solution of the Helmholtz equation satisfying the Sommerfeld
condition at infinity can be represented as a single layer potential which
satisfies (\ref{redund1}).

When $u=\overrightarrow{u}$, the vector part of (\ref{radHelm}) gives us (%
\ref{Som2}) and the scalar part%
\begin{equation*}
\left\langle \frac{x}{\left| x\right| },\limfunc{rot}\overrightarrow{u}%
(x)\right\rangle =o(\frac{1}{\left| x\right| })
\end{equation*}%
is again a simple consequence from the integral representation of $%
\overrightarrow{u}$ (see \cite[Sect. 4.2]{CK1} or \cite[p. 120]{KSbook}).

Thus (\ref{radHelm}) in special cases reduces to (\ref{Som1}) and (\ref{Som2}%
) and in general represents the radiation condition at infinity for the
quaternionic Helmholtz equation.

Note that (\ref{rad+}) and (\ref{rad-}) follow from (\ref{radHelm})
immediately if one assumes that $u\in \ker D_{\alpha }$ or $u\in \ker
D_{-\alpha }$ respectively.

We will need the following operators%
\begin{equation*}
S_{\alpha }f(x)=-2\int_{\Gamma }\mathcal{K}_{\alpha }(x-y)\overrightarrow{n}%
(y)f(y)d\Gamma _{y},\quad x\in \Gamma ,
\end{equation*}%
\begin{equation*}
P_{\alpha }=\frac{1}{2}(I+S_{\alpha })\quad \text{and\quad }Q_{\alpha }=%
\frac{1}{2}(I-S_{\alpha })
\end{equation*}%
defined for example on H\"{o}lder functions in the sense of the Cauchy
principal value. It is well known that $S_{\alpha }$ is a singular integral
operator of the Calderon-Zygmund type (see \cite[Sect. 2.5]{GS1}). This
implies the boundedness of the operators $P_{\alpha }$, $Q_{\alpha }$ and $%
S_{\alpha }$ in Sobolev spaces $H^{s}(\Gamma )$ for all real $s$.

The following important properties of the operators $P_{\alpha }$, $%
Q_{\alpha }$ and $S_{\alpha }$ will be widely used in this work.

\begin{theorem}
\label{PSf} Let $f\in L_{2}(\Gamma )$. Then for an almost every point $\tau
\in \Gamma $ there exist the following nontangential limits 
\begin{equation}
\lim_{\Omega ^{\pm }\ni x\rightarrow \tau \in \Gamma }K_{\alpha
}[f](x)=K_{\alpha }[f]^{\pm }(\tau ),  \label{t1}
\end{equation}%
and the following formulas hold 
\begin{equation}
K_{\alpha }[f]^{+}(\tau )=P_{\alpha }[f](\tau )=\frac{1}{2}(I+S_{\alpha
})f(\tau ),\qquad K_{\alpha }[f]^{-}(\tau )=-Q_{\alpha }[f](\tau )=-\frac{1}{%
2}(I-S_{\alpha })f(\tau ).  \label{t2}
\end{equation}
\end{theorem}

Let now $\Gamma $ be a sufficiently smooth surface in order that the Sobolev
space $H^{s}(\Gamma )$ for a given $s$ be defined.

\begin{remark}
\label{Remark} Since $H^{s}(\Gamma )\subset L_{2}(\Gamma )$ for $s\geq 0$
and $S_{\alpha }$ is bounded in $H^{s}(\Gamma )$, equalities (\ref{t1}) and (%
\ref{t2}) hold for $f\in H^{s}(\Gamma )$, $s\geq 0$.
\end{remark}

\begin{corollary}
\begin{enumerate}
\item The following equalities hold on $H^{s}(\Gamma )$, $s\geq 0$ 
\begin{equation*}
S_{\alpha }^{2}=I,\quad P_{\alpha }^{2}=P_{\alpha },\quad Q_{\alpha
}^{2}=Q_{\alpha },\quad P_{\alpha }Q_{\alpha }=Q_{\alpha }P_{\alpha }=0.
\end{equation*}%
That is $P_{\alpha }$ and $Q_{\alpha }$ are orthogonal projection operators
on $H^{s}(\Gamma )$, $s\geq 0$.

\item In order for $f\in H^{s}(\Gamma )$, $s\geq 0$ to be a boundary value
of a function $F\in \ker D_{\alpha }(\Omega ^{+})$, the following condition
is necessary and sufficient: 
\begin{equation}
f\in \func{im}P_{\alpha }(H^{s}(\Gamma )).  \label{critP}
\end{equation}

\item In order for $f\in H^{s}(\Gamma )$, $s\geq 0$ to be a boundary value
of a function $F\in \ker D_{\alpha }(\Omega ^{-})$, satisfying (\ref{rad+})
at infinity, the following condition is necessary and sufficient: 
\begin{equation}
f\in \func{im}Q_{\alpha }(H^{s}(\Gamma )).  \label{critQ}
\end{equation}
\end{enumerate}
\end{corollary}

The proof of these facts in $L_{2}(\Gamma )$ can be found in \cite[Chapter 5]%
{KSbook}. By Theorem \ref{PSf} and Remark \ref{Remark} these assertions also
hold in the space $H^{s}(\Gamma )$, $s\geq 0$.

Needless to say that the same facts are valid for $D_{-\alpha }$.

\section{Complete systems of fundamental solutions of the Helmholtz operator%
\label{points}}

Let $\Gamma $ be a closed surface in $\mathbb{R}^{3}$ which is a boundary of
a bounded domain $\Omega ^{+}$ and of an unbounded domain $\Omega ^{-}=%
\mathbb{R}^{3}\setminus \overline{\Omega ^{+}}$. By $\Gamma ^{-}$ we denote
a closed surface enclosed in $\Omega ^{+}$ and enclosing the domain $V$ and
by $\Gamma ^{+}$ a closed surface enclosing $\overline{\Omega ^{+}}$ as
shown in Fig. \ref{Gammi}.\FRAME{ftbpF}{3.7351in}{3.0545in}{0pt}{}{\Qlb{Gammi%
}}{fig1.bmp}{\special{language "Scientific Word";type
"GRAPHIC";maintain-aspect-ratio TRUE;display "USEDEF";valid_file "F";width
3.7351in;height 3.0545in;depth 0pt;original-width 3.6876in;original-height
3.0104in;cropleft "0";croptop "1";cropright "1";cropbottom "0";filename
'Fig1.bmp';file-properties "XNPEU";}} By $\left\{ y_{n}^{-}\right\}
_{n=1}^{\infty }$ we denote a set of points distributed on $\Gamma ^{-}$ and
dense on $\Gamma ^{-}$, and by $\left\{ y_{n}^{+}\right\} _{n=1}^{\infty }$
a set of points distributed on $\Gamma ^{+}$ and dense on $\Gamma ^{+}$. To
each of these sets a system of fundamental solutions $\left\{ \theta
_{\alpha }(x-y_{n}^{-})\right\} _{n=1}^{\infty }$ or $\left\{ \theta
_{\alpha }(x-y_{n}^{+})\right\} _{n=1}^{\infty }$ is related. Denote $\theta
_{\alpha ,n}^{-}(x)=\theta _{\alpha }(x-y_{n}^{-})$ and $\theta _{\alpha
,n}^{+}(x)=\theta _{\alpha }(x-y_{n}^{+})$. Singularities of functions of
the first system are distributed on the interior surface $\Gamma ^{-}$ and
consequently every such function is a solution of the Helmholtz equation in $%
\Omega ^{-}$ satisfying the Sommerfeld radiation condition at infinity.
Functions from the second system have their singularities on the exterior
surface $\Gamma ^{+}$ and solve the Helmholtz equation in $\Omega ^{+}$.

We start with the following theorem due to V. Kupradze \cite{Kupr}, the
proof can be found, for example, in \cite[p. 51]{Eremin}.

\begin{theorem}
\label{DEW} Let $\Gamma $ be a closed surface of class $C^{2}$. Then the
system of functions $\left\{ \theta _{\alpha }(x-y_{n}^{+})\right\}
_{n=1}^{\infty }$ is complete in $L_{2}(\Gamma )$. Assume additionally that $%
\alpha ^{2}$ is not an eigenvalue of the Dirichlet problem in $V$. Then the
system of functions $\left\{ \theta _{\alpha }(x-y_{n}^{-})\right\}
_{n=1}^{\infty }$ is complete in $L_{2}(\Gamma )$ also.
\end{theorem}

Our aim is to obtain a similar result for the Sobolev spaces $H^{s}(\Gamma )$%
. This will require a sequence of steps. We will show first that these
systems of functions are complete in $L_{2}(\Omega )\cap \ker (\Delta
+\alpha ^{2})$. Then this result will be extended to $H^{s}(\Omega )\cap
\ker (\Delta +\alpha ^{2})$. Finally as the space $H^{s}(\Gamma )$ can be
considered as a space of traces of corresponding solutions of the Helmholtz
equation we will be able to prove the completeness of our systems of
fundamental solutions for the Helmholtz operator in this space. Let us
consider first the case of a bounded domain $\Omega ^{+}$ and then of an
unbounded domain $\Omega ^{-}$.

\subsection{Interior domain}

\begin{theorem}
\label{Th9}Let $\Omega ^{+}$ be a bounded domain in $\mathbb{R}^{3}$ with a
Liapunov boundary $\Gamma $. \ The system of functions $\left\{ \theta
_{\alpha ,n}^{+}\right\} _{n=1}^{\infty }$ is complete in $L_{2}(\Omega
^{+})\cap \ker (\Delta +\alpha ^{2})$.
\end{theorem}

\begin{proof}
We consider the operator $\Delta +\alpha ^{2}$ as an unbounded operator in $%
L_{2}(\Omega ^{+})$ with the domain $H^{2}(\Omega ^{+})$. This operator is
closed and the set $L_{2}(\Omega ^{+})\cap \ker (\Delta +\alpha ^{2})$ is a
subspace. Thus it is sufficient to prove that the system $\left\{ \theta
_{\alpha ,n}^{+}\right\} _{n=1}^{\infty }$ is closed in $L_{2}(\Omega
^{+})\cap \ker (\Delta +\alpha ^{2})$. Assume that there exists a
non-trivial function $f\in L_{2}(\Omega ^{+})\cap \ker (\Delta +\alpha ^{2})$
with the property%
\begin{equation*}
\left\langle \theta _{\alpha ,n}^{+},f\right\rangle _{L_{2}(\Omega ^{+})}=0%
\text{\quad for all }n\in \mathbb{N},
\end{equation*}%
or in explicit form:%
\begin{equation*}
\int_{\Omega ^{+}}\theta _{\alpha }(x-y_{n}^{+})f^{\ast }(x)dx=0\text{\quad
for all }n\in \mathbb{N},
\end{equation*}%
where ``*'' stands for the usual complex conjugation. Denote%
\begin{equation*}
V_{\Omega ^{+}}f(y)=\int_{\Omega ^{+}}\theta _{\alpha }(x-y)f(x)dx.
\end{equation*}%
We have that $V_{\Omega ^{+}}f^{\ast }(y_{n}^{+})=0$ for all $n\in \mathbb{N}
$. These equalities and the continuity of $V_{\Omega ^{+}}f^{\ast }$ imply
the equality $V_{\Omega ^{+}}f^{\ast }=0$ on $\Gamma ^{+}$.

The function $V_{\Omega ^{+}}f^{\ast }$ satisfies the Helmholtz equation in $%
\Omega ^{-}$ and fulfills the Sommerfeld radiation condition at infinity.
Consequently $V_{\Omega ^{+}}f^{\ast }\equiv 0$ in $\Omega ^{-}$. Moreover,
all the derivatives of $V_{\Omega ^{+}}f^{\ast }$ in $\Omega ^{-}$ are equal
to zero. Thus we obtain that the function $V_{\Omega ^{+}}f^{\ast }$ and all
its derivatives are equal to zero on $\Gamma $. Taking this into account and
using the fact that $V_{\Omega ^{+}}f^{\ast }\in \ker (\Delta +\alpha ^{\ast
2})(\Delta +\alpha ^{2})$ in $\Omega ^{+}$ due to the uniqueness of
continuation for the null solutions of this elliptic operator \cite[Theorem
6.14]{Mizohata} we obtain that $V_{\Omega ^{+}}f^{\ast }\equiv 0$ in $\Omega
^{+}$ and hence $f\equiv 0$ in $\Omega ^{+}$.
\end{proof}

\begin{theorem}
\label{Th10}Under the conditions of Theorem \ref{Th9} the system of
functions $\left\{ \theta _{\alpha ,n}^{+}\right\} _{n=1}^{\infty }$ is
complete in $H^{s}(\Omega ^{+})\cap \ker (\Delta +\alpha ^{2})$, $s\geq 0$.
\end{theorem}

\begin{proof}
Here we use a quite general fact proved by N. Tarkhanov (for a general
elliptic system) \cite[Sect. 8.1]{Tarkhanov} that a function from $%
H^{s}(\Omega ^{+})\cap \ker (\Delta +\alpha ^{2})$ belongs to the closure of
the subspace $\func{sol}(\Omega ^{+})$ in $H^{s}(\Omega ^{+})$ consisting of
all $C^{\infty }$ solutions of the Helmholtz equation in a neighborhood of $%
\overline{\Omega ^{+}}$. That is for any function $f\in H^{s}(\Omega
^{+})\cap \ker (\Delta +\alpha ^{2})$ and for any $\varepsilon >0$ we can
find such a function $\ f_{0}\in \ker (\Delta +\alpha ^{2})$ in $\widetilde{%
\Omega }^{+}$, where $\overline{\Omega ^{+}}\subset \widetilde{\Omega }^{+}$
that $\left\| f-f_{0}\right\| _{H^{s}(\Omega ^{+})}<\varepsilon /2$. The
domain $\widetilde{\Omega }^{+}$ can be chosen enclosed by $\Gamma ^{+}$.

For all solutions $u$ of the Helmholtz equation in $\widetilde{\Omega }^{+}$
we have the following estimate (see, e.g., \cite[Theorem 11.1]{Taylor})%
\begin{equation*}
\left\| u\right\| _{H^{s}(\Omega ^{+})}\leq C\left\| u\right\| _{L_{2}(%
\widetilde{\Omega }^{+})},
\end{equation*}%
where the constant $C$ does not depend on $u$. Due to Theorem \ref{Th9} for
any $\varepsilon _{1}>0$ the function $f_{0}$ can be approximated by a
linear combination $f_{N}=\sum_{n=1}^{N}a_{n}\theta _{\alpha ,n}^{+}$ in $%
\widetilde{\Omega }^{+}$ in such a way that%
\begin{equation*}
\left\| f_{0}-f_{N}\right\| _{L_{2}(\widetilde{\Omega }^{+})}<\varepsilon
_{1}.
\end{equation*}%
Choose $\varepsilon _{1}=\varepsilon /(2C)$ and consider%
\begin{equation*}
\left\| f-f_{N}\right\| _{H^{s}(\Omega ^{+})}=\left\|
f-f_{0}+f_{0}-f_{N}\right\| _{H^{s}(\Omega ^{+})}\leq \left\|
f-f_{0}\right\| _{H^{s}(\Omega ^{+})}+\left\| f_{0}-f_{N}\right\|
_{H^{s}(\Omega ^{+})}<
\end{equation*}%
\begin{equation*}
\frac{\varepsilon }{2}+C\left\| f_{0}-f_{N}\right\| _{L_{2}(\widetilde{%
\Omega }^{+})}<\varepsilon
\end{equation*}
\end{proof}

\begin{theorem}
\label{Th11} Let $\Gamma $ be a sufficiently smooth (the space $H^{s}(\Gamma
)$ is defined) closed surface. The system of functions $\left\{ \theta
_{\alpha ,n}^{+}\right\} _{n=1}^{\infty }$ is complete in $H^{s}(\Gamma )$, $%
s\in \mathbb{R}$.
\end{theorem}

\begin{proof}
For $s\leq 0$ the result follows from Theorem \ref{Th9}. Let $s>0$, given $%
\varepsilon >0$ then for any $u\in H^{s}(\Gamma )$ there exists (probably
not unique) a solution of the Dirichlet problem $(\Delta +\alpha ^{2})U=0$
in $\Omega ^{+}$ and $U\left| _{\Gamma }\right. =u$, a function $U\in
H^{s+1/2}(\Omega ^{+})$. Due to Theorem \ref{Th10} for any $\varepsilon
_{1}>0$ we can approximate it by a linear combination $U_{N}=%
\sum_{n=1}^{N}a_{n}\theta _{\alpha ,n}^{+}$ in such a way that%
\begin{equation*}
\left\| U-U_{N}\right\| _{H^{s+1/2}(\Omega ^{+})}<\varepsilon _{1}.
\end{equation*}%
Using the continuity of the trace operator $\gamma $ we obtain that%
\begin{equation*}
\left\| u-u_{N}\right\| _{H^{s}(\Gamma )}=\left\| \gamma (U-U_{N})\right\|
_{H^{s}(\Gamma )}\leq C\left\| U-U_{N}\right\| _{H^{s+1/2}(\Omega
^{+})}<C\varepsilon _{1}.
\end{equation*}%
Choosing $\varepsilon _{1}=\varepsilon /C$ we finish the proof.
\end{proof}

\begin{remark}
This theorem was proved for scalar functions from $H^{s}(\Gamma )$.
Nevertheless it is obviously valid also for $\mathbb{H}(\mathbb{C})$-valued
functions from $H^{s}(\Gamma )$ which in this case are approximated by
linear combinations $\sum_{n=1}^{N}c_{n}^{+}\theta _{\alpha ,n}^{+}$ where $%
c_{n}^{+}$ are complex quaternions.
\end{remark}

\bigskip

\subsection{Exterior domain}

Let $B_{R}$ be an arbitrary ball with a sufficiently large radius $R$ such
that $\Omega ^{+}\subset B_{R}$. Denote $\Omega _{R}^{-}=\Omega ^{-}\cap
B_{R}$. Thus $\Omega _{R}^{-}$ is a domain in $\mathbb{R}^{3}$ with a
boundary consisting of $\Gamma $ and of the sphere $\partial B_{R}$.

\begin{theorem}
\label{Th12}Let $\Gamma $ be a closed Liapunov surface and $\alpha ^{2}$ be
not an eigenvalue of the Dirichlet problem in $V$. The system of functions $%
\left\{ \theta _{\alpha ,n}^{-}\right\} _{n=1}^{\infty }$ is complete in $%
L_{2}(\Omega _{R}^{-})\cap \ker (\Delta +\alpha ^{2})$.
\end{theorem}

\begin{proof}
Assume that there exists a non-trivial function $f\in L_{2}(\Omega
_{R}^{-})\cap \ker (\Delta +\alpha ^{2})$ with the property%
\begin{equation*}
\left\langle \theta _{\alpha ,n}^{-},f\right\rangle _{L_{2}(\Omega
_{R}^{-})}=0\text{\quad for all }n\in \mathbb{N},
\end{equation*}%
or in explicit form:%
\begin{equation*}
\int_{\Omega _{R}^{-}}\theta _{\alpha }(x-y_{n}^{-})f^{\ast }(x)dx=0\text{%
\quad for all }n\in \mathbb{N}.
\end{equation*}%
Denote%
\begin{equation*}
V_{\Omega _{R}^{-}}f(y)=\int_{\Omega _{R}^{-}}\theta _{\alpha }(x-y)f(x)dx.
\end{equation*}%
We have that $V_{\Omega _{R}^{-}}f^{\ast }(y_{n}^{-})=0$ for all $n\in 
\mathbb{N}$ and hence $V_{\Omega _{R}^{-}}f^{\ast }=0$ on $\Gamma ^{-}$.

The function $V_{\Omega _{R}^{-}}f^{\ast }$ satisfies the Helmholtz equation
in $\Omega ^{+}$. It is equal to zero in $V$ due to the uniqueness of a
solution of the Dirichlet problem in $V$ and it is zero with all its
derivatives in $\overline{\Omega ^{+}}$ due to the uniqueness of
continuation for the solutions of the Helmholtz equation. Moreover in $%
\overline{\Omega ^{+}}\cup \Omega _{R}^{-}$ the function $V_{\Omega
_{R}^{-}}f^{\ast }$ belongs to $\ker (\Delta +\alpha ^{\ast 2})(\Delta
+\alpha ^{2})$. Thus due to the uniqueness of continuation for the solutions
of this elliptic operator we obtain that $V_{\Omega _{R}^{-}}f^{\ast }\equiv
0$ in $\Omega _{R}^{-}$ and hence $f\equiv 0$ in $\Omega _{R}^{-}$.
\end{proof}

\begin{theorem}
Under the conditions of the previous theorem the system of functions $%
\left\{ \theta _{\alpha ,n}^{-}\right\} _{n=1}^{\infty }$ is complete in $%
H^{s}(\Omega _{R}^{-})\cap \ker (\Delta +\alpha ^{2})$, $s\geq 0$.
\end{theorem}

\begin{proof}
First, for a given $\varepsilon >0$ we choose such a function $f_{0}\in \ker
(\Delta +\alpha ^{2})$ in $\widetilde{\Omega }_{R}^{-}$ that $\left\|
f-f_{0}\right\| _{H^{s}(\Omega _{R}^{-})}<\varepsilon /2$, where $\widetilde{%
\Omega }_{R}^{-}$ is a domain containing $\overline{\Omega _{R}^{-}}$ and
such that $\Gamma ^{-}\cap \widetilde{\Omega }_{R}^{-}=\emptyset $. For all
solutions $u$ of the Helmholtz equation in $\widetilde{\Omega }_{R}^{-}$ we
have the following estimate 
\begin{equation*}
\left\| u\right\| _{H^{s}(\Omega _{R}^{-})}\leq C\left\| u\right\| _{L_{2}(%
\widetilde{\Omega }_{R}^{-})},
\end{equation*}%
where the constant $C$ does not depend on $u$.

The proof finishes by analogy with that of Theorem \ref{Th10}.
\end{proof}

\begin{theorem}
\label{Th14}Let $\Gamma $ be a sufficiently smooth closed surface and $%
\alpha ^{2}$ be not an eigenvalue of the Dirichlet problem in $V$. The
system of functions $\left\{ \theta _{\alpha ,n}^{-}\right\} _{n=1}^{\infty
} $ is complete in $H^{s}(\Gamma )$, $s\in \mathbb{R}$.
\end{theorem}

The proof is completely analogous to that of Theorem \ref{Th11}.

\section{\protect\bigskip Extensions into exterior domains\label{ext}}

Let us consider the exterior Dirichlet problem for the Helmholtz equation%
\begin{eqnarray}
(\Delta +\alpha ^{2})U &=&0\qquad \text{in }\Omega ^{-},  \notag \\
&&  \label{Dirpr} \\
U|_{\Gamma } &=&u,  \notag
\end{eqnarray}%
and $U$ satisfies the Sommerfeld radiation condition (\ref{Som1}) at
infinity. For $u\in H^{s}(\Gamma )$, $s>0$ it is known that the solution of
this problem exists, is unique and belongs to a weighted Sobolev space in $%
\Omega ^{-}$ (see \cite[Sect. 2.6]{Nedelec}). For our purposes the important
will be the fact that the solution belongs to $H^{s+1/2}(\Omega _{R}^{-})$
where $\Omega _{R}^{-}$ is an intersection of $\Omega ^{-}$ with a ball $%
B_{R}$ of radius $R$ chosen large enough to enclose the interior domain $%
\Omega ^{+}$. We denote by $H_{loc}^{s+1/2}(\Omega ^{-})$ the union of all
such $H_{{}}^{s+1/2}(\Omega _{R}^{-})$.

The same will be valid if in (\ref{Dirpr}) we assume the functions $u$ and $%
U $ to be $\mathbb{H}(\mathbb{C})$-valued and each Cartesian component of $U$
satisfy (\ref{Som1}) or which is equivalent the whole function $U$ satisfy (%
\ref{radHelm}). The operator transforming $u$ into $U$ we denote by $\Lambda 
$, and as we have just seen\ $\Lambda $ acts from $H^{s}(\Gamma )\ $to $%
H_{loc}^{s+1/2}(\Omega ^{-})$.

The operators $\Pi _{\pm \alpha }$ introduced above act obviously from $%
H_{loc}^{s}(\Omega ^{-})$ to $H_{loc}^{s-1}(\Omega ^{-})$. Consider a
function $\Pi _{\alpha }\Lambda u$. For $u\in H^{s}(\Gamma )$, $s>1$ it will
belong to $H_{loc}^{s-1/2}(\Omega ^{-})$ and its trace\ (see, e.g., \cite[p.
50]{Nedelec}) $\gamma \Pi _{\alpha }\Lambda u\in H^{s-1}(\Gamma )$. As the
operators $Q_{\pm \alpha }$ are bounded in $H^{s}(\Gamma )$, we can
introduce two new operators $\widetilde{Q}_{\alpha }$ and $\widetilde{Q}%
_{-\alpha }$ as follows%
\begin{equation*}
\widetilde{Q}_{\pm \alpha }=Q_{\pm \alpha }\gamma \Pi _{\pm \alpha }\Lambda
:H^{s}(\Gamma )\rightarrow H^{s-1}(\Gamma ),\qquad s>1.
\end{equation*}

\begin{proposition}
\label{Q+Q}Let an $\mathbb{H}(\mathbb{C})$-valued function $u\ $belongs to $%
H^{s}(\Gamma )$, $s>1$. Then 
\begin{equation}
u=\widetilde{Q}_{\alpha }u+\widetilde{Q}_{-\alpha }u.  \label{uQ}
\end{equation}
\end{proposition}

\begin{proof}
Consider $U=\Lambda u$. We have 
\begin{equation}
U=\Pi _{\alpha }U+\Pi _{-\alpha }U  \label{UP}
\end{equation}%
\begin{equation*}
=-(K_{\alpha }\gamma \Pi _{\alpha }U+K_{-\alpha }\gamma \Pi _{-\alpha }U),
\end{equation*}%
for any point $x\in \Omega ^{-}$. Taking the limit of this equality when $x$
tends to the boundary and using Theorem \ref{PSf} we obtain that 
\begin{equation*}
u=Q_{\alpha }\gamma \Pi _{\alpha }\Lambda u+Q_{-\alpha }\gamma \Pi _{-\alpha
}\Lambda u=\widetilde{Q}_{\alpha }u+\widetilde{Q}_{-\alpha }u.
\end{equation*}
\end{proof}

\begin{remark}
As we have seen the operators $\widetilde{Q}_{\pm \alpha }$ act from $%
H^{s}(\Gamma )$ to $H^{s-1}(\Gamma )$, $s>1$, so equality (\ref{uQ}) can
appear a little bit surprising. Nevertheless this is a reflection of the
corresponding fact inside the domain $\Omega ^{-}$ (\ref{UP}), where the
differential operators $\Pi _{\pm \alpha }$ also reduce the smoothness of a
function but the derivatives in (\ref{UP}) are cancelled.
\end{remark}

\begin{proposition}
\label{Qus} Let $f\in \func{im}Q_{\alpha }(H^{s}(\Gamma ))$, $s>0$. Then $%
Q_{\alpha }f=\widetilde{Q}_{\alpha }f$.
\end{proposition}

\begin{proof}
Let $f\in \func{im}Q_{\alpha }$, that is $f=Q_{\alpha }f$. We have that the
function $\Lambda f$ satisfies equation (\ref{D+}) in $\Omega ^{-}$ and
belongs to $H_{loc}^{s+1/2}(\Omega ^{-})$. Moreover due to the uniqueness of
the solution of the Dirichlet problem for the Helmholtz operator in $\Omega
^{-}$ we obtain that $\Pi _{\alpha }\Lambda f=\Lambda f$. Thus%
\begin{equation*}
\Lambda f=-K_{\alpha }\gamma \Pi _{\alpha }\Lambda f
\end{equation*}%
which on the boundary due to Theorem \ref{PSf} gives us that $f=\widetilde{Q}%
_{\alpha }f$.
\end{proof}

Let us introduce the following systems of functions 
\begin{equation}
\left\{ \mathcal{K}_{\alpha ,n}^{\pm }(x)=(-D+\alpha )\theta _{\alpha
}(x-y_{n}^{\pm })\right\} _{n=1}^{\infty }  \label{Kfund+}
\end{equation}%
and 
\begin{equation}
\left\{ \mathcal{K}_{-\alpha ,n}^{\pm }(x)=-(D+\alpha )\theta _{\alpha
}(x-y_{n}^{\pm })\right\} _{n=1}^{\infty }  \label{Kfund-}
\end{equation}%
where the sets of points $\left\{ y_{n}^{+}\right\} _{n=1}^{\infty }$ and $%
\left\{ y_{n}^{-}\right\} _{n=1}^{\infty }$ are defined as in Section \ref%
{points}. We are ready to prove one of the central facts of this work.

\begin{theorem}
\label{complext}Let $\alpha ^{2}$ be not an eigenvalue of the Dirichlet
problem in $V$. Then the systems of functions $\left\{ \mathcal{K}_{\pm
\alpha ,n}^{-}\right\} _{n=1}^{\infty }$ are complete in $\func{im}Q_{\pm
\alpha }(H^{s}(\Gamma ))$, $s>1$ respectively by the norm of $H^{s-1}(\Gamma
)$.
\end{theorem}

\begin{proof}
Let us consider the system $\left\{ \mathcal{K}_{\alpha ,n}^{-}\right\}
_{n=1}^{\infty }$. Due to Proposition \ref{Qus} any function $f\in \func{im}%
Q_{\alpha }(H^{s}(\Gamma ))$ can be represented as follows%
\begin{equation*}
f=\widetilde{Q}_{\alpha }f.
\end{equation*}%
Due to Theorem \ref{Th14} for any $\epsilon >0$ there exists such a linear
combination%
\begin{equation*}
f_{N}=\sum_{j=1}^{N}a_{j}\theta _{\alpha ,j}^{-}
\end{equation*}%
that 
\begin{equation*}
\left\| f-f_{N}\right\| _{H^{s}(\Gamma )}<\epsilon .\text{ }
\end{equation*}%
Here $a_{j}$ are constant complex quaternions. Due to the boundedness of $%
\widetilde{Q}_{\alpha }$ we have%
\begin{equation*}
\left\| f-\widetilde{Q}_{\alpha }f_{N}\right\| _{H^{s-1}(\Gamma )}=\left\| 
\widetilde{Q}_{\alpha }f-\widetilde{Q}_{\alpha }f_{N}\right\|
_{H^{s-1}(\Gamma )}\leq C\left\| f-f_{N}\right\| _{H^{s}(\Gamma )}<\epsilon
\end{equation*}%
where $C$ is a positive constant. Thus the function $\widetilde{Q}_{\alpha
}f_{N}$ approximates $f$ in the norm of $H^{s-1}(\Gamma )$. Consider%
\begin{equation*}
\widetilde{Q}_{\alpha }f_{N}=Q_{\alpha }\gamma \Pi _{\alpha }\Lambda
\sum_{j=1}^{N}a_{j}\theta _{\alpha ,j}^{-}.
\end{equation*}%
It is obvious that the extension $\Lambda \theta _{\alpha ,j}^{-}$ coincides
with the values of $\theta _{\alpha }(x-y_{j}^{-})$ for all $x\in \Omega
^{-} $. We obtain%
\begin{equation*}
\widetilde{Q}_{\alpha }f_{N}(x)=\sum_{j=1}^{N}Q_{\alpha }\gamma \Pi _{\alpha
}(\theta _{\alpha }(x-y_{j}^{-}))a_{j}=\frac{1}{2\alpha }\sum_{j=1}^{N}Q_{%
\alpha }\gamma \mathcal{K}_{\alpha ,j}^{-}(x)a_{j}=\frac{1}{2\alpha }%
\sum_{j=1}^{N}\mathcal{K}_{\alpha ,j}^{-}(x)a_{j}.
\end{equation*}
\end{proof}

\section{Extensions into interior domains}

The results of this section and their proofs are similar to those of Section %
\ref{ext} and we present them more briefly. Here a new and natural
assumption will be that $\alpha ^{2}$ is not an eigenvalue of the Dirichlet
problem in $\Omega ^{+}$. Then for each $\mathbb{H}(\mathbb{C})$-valued
function $u\in H^{s}(\Gamma )$, $s>0$ there exists its unique Helmholtz
extension, an $\mathbb{H}(\mathbb{C})$-valued function $U$ satisfying the
Helmholtz equation (\ref{Helmeq}) in $\Omega ^{+}$ and coinciding with $u$
on the boundary. As before the operator transforming $u$ into $U$ we denote
by $\Lambda $. By analogy with the operators $\widetilde{Q}_{\pm \alpha }$
we introduce the operators

\begin{equation*}
\widetilde{P}_{\pm \alpha }=P_{\pm \alpha }\gamma \Pi _{\pm \alpha }\Lambda
:H^{s}(\Gamma )\rightarrow H^{s-1}(\Gamma ),\qquad s>1.
\end{equation*}

\begin{proposition}
Let $\alpha ^{2}$ be not an eigenvalue of the Dirichlet problem in $\Omega
^{+}$ and an $\mathbb{H}(\mathbb{C})$-valued function $u\ $belong to $%
H^{s}(\Gamma )$, $s>1$. Then 
\begin{equation*}
u=\widetilde{P}_{\alpha }u+\widetilde{P}_{-\alpha }u.
\end{equation*}
\end{proposition}

The proof is analogous to that of Proposition \ref{Q+Q}.

\begin{proposition}
Let $\alpha ^{2}$ be not an eigenvalue of the Dirichlet problem in $\Omega
^{+}$ and an $\mathbb{H}(\mathbb{C})$-valued function $f$ belong to $\func{im%
}P_{\alpha }(H^{s}(\Gamma ))$, $s>0$. Then $P_{\alpha }f=$ $\widetilde{P}%
_{\alpha }f$.
\end{proposition}

The proof is analogous to that of Proposition \ref{Qus}.

Finally, by analogy with Theorem \ref{complext} the following statement is
proved

\begin{theorem}
Let $\alpha ^{2}$ be not an eigenvalue of the Dirichlet problem in $\Omega
^{+}$. Then the systems of functions $\left\{ \mathcal{K}_{\pm \alpha
,n}^{+}\right\} _{n=1}^{\infty }$ are complete in $\func{im}P_{\pm \alpha
}(H^{s}(\Gamma ))$, $s>1$ respectively by the norm of $H^{s-1}(\Gamma )$.
\end{theorem}

\begin{remark}
For $\alpha =0$ a similar result can be found in \cite[p. 284]{GS2} (see
also references therein). Unfortunately the scheme of the proof proposed in
that work is not applicable for complex quaternion valued functions due to
the difficulty of introduction of an $L_{2}$ space which would correspond to
the complex quaternionic multiplication.
\end{remark}

\section{Complete systems for Maxwell's equations}

As we will see our approach works not only for homogeneous, isotropic,
achiral media but also for chiral media. This last case is more general.
When the chirality measure of a medium $\beta $ is equal to zero we obtain
the nonchiral or achiral situation. This is why we show our results for the
case of a chiral medium, transition to a nonchiral case is quite easy.

For the sake of simplicity we consider a sourceless situation. Then
Maxwell's equations for time-harmonic electromagnetic fields in a chiral
medium have the form (see, e.g., \cite{LVV, Lindell})

\begin{equation}
\limfunc{div}\widetilde{E}\left( x\right) =\limfunc{div}\widetilde{H}\left(
x\right) =0,  \label{Mc1}
\end{equation}

\begin{equation}
\func{rot}\widetilde{E}\left( x\right) =i\omega\widetilde{B}\left( x\right) ,
\label{Mc2}
\end{equation}

\begin{equation}
\func{rot}\widetilde{H}\left( x\right) =-i\omega \widetilde{D}\left(
x\right) ,  \label{Mc3}
\end{equation}%
with the constitutive relations \cite{LVV}

\begin{equation}
\widetilde{D}=\varepsilon\left( \widetilde{E}\left( x\right) +\beta \func{rot%
}\widetilde{E}\left( x\right) \right) ,  \label{M5}
\end{equation}

\begin{equation}
\widetilde{B}=\mu \left( \widetilde{H}\left( x\right) +\beta \func{rot}%
\widetilde{H}\left( x\right) \right) ,  \label{M6}
\end{equation}%
where $\omega $ is the frequency, $\varepsilon $ and $\mu $ are\ complex
permittivity and permeability of a medium and $\beta $ \ is its chirality
measure.

The Maxwell equations (\ref{Mc1})-(\ref{Mc3}) can be also written as follows

\begin{equation}
\func{rot}\widetilde{E}\left( x\right) =i\omega\mu\left( \widetilde{H}\left(
x\right) +\beta\func{rot}\widetilde{H}\left( x\right) \right) ,  \label{M7}
\end{equation}

\begin{equation}
\func{rot}\widetilde{H}\left( x\right) =-i\omega \varepsilon \left( 
\widetilde{E}\left( x\right) +\beta \func{rot}\widetilde{E}\left( x\right)
\right) .  \label{M8}
\end{equation}%
Introducing the notations

\begin{equation}
\widetilde{E}\left( x\right) =-\sqrt{\mu}\cdot\overrightarrow{E}\left(
x\right) ,  \label{M9}
\end{equation}

\begin{equation}
\widetilde{H}\left( x\right) =\sqrt{\varepsilon }\cdot \overrightarrow{H}%
\left( x\right) ,  \label{M10}
\end{equation}%
we obtain the equations

\begin{equation}
\func{rot}\overrightarrow{E}\left( x\right) =-i\alpha\left( \overrightarrow{H%
}\left( x\right) +\beta\func{rot}\overrightarrow {H}\left( x\right) \right)
\label{M12}
\end{equation}
and

\begin{equation}
\func{rot}\overrightarrow{H}\left( x\right) =i\alpha \left( \overrightarrow{E%
}\left( x\right) +\beta \func{rot}\overrightarrow{E}\left( x\right) \right) ,
\label{M13}
\end{equation}%
where as before $\alpha =\omega \sqrt{\varepsilon \mu }$ and in the case of $%
\beta =0$, $\alpha $ is the wave number. When $\beta $ is different from
zero, as it will be seen below, $\alpha $ does not have the same physical
meaning. There appear two wave numbers instead, $\alpha _{1}$ and $\alpha
_{2}$.

Let us consider the following purely vectorial biquaternionic functions:

\begin{equation}
\overrightarrow{\varphi}(x)=\overrightarrow{E}\left( x\right) +i%
\overrightarrow{H}\left( x\right)  \label{M14}
\end{equation}
and

\begin{equation}
\overrightarrow{\psi}(x)=\overrightarrow{E}\left( x\right) -i%
\overrightarrow {H}\left( x\right) .  \label{M15}
\end{equation}
We have that

\begin{equation}
D\overrightarrow{\varphi }\left( x\right) =\func{rot}\overrightarrow{E}%
\left( x\right) +i\func{rot}\overrightarrow{H}\left( x\right) .  \label{M21}
\end{equation}%
Using (\ref{M12}) and (\ref{M13}) we obtain%
\begin{align*}
D\overrightarrow{\varphi }\left( x\right) & =-(i\alpha \overrightarrow{H}%
\left( x\right) +\alpha \overrightarrow{E}\left( x\right) )- \\
& \\
& \alpha \beta \left( D\overrightarrow{E}\left( x\right) +iD\overrightarrow{H%
}\left( x\right) \right) .
\end{align*}%
That is,

\begin{equation*}
D\overrightarrow{\varphi }\left( x\right) =-\alpha \overrightarrow{\varphi }%
\left( x\right) -\alpha \beta D\overrightarrow{\varphi }\left( x\right) .
\end{equation*}%
Thus the complex quaternionic function $\overrightarrow{\varphi }$ satisfies
the following equation%
\begin{equation}
\left( D+\frac{\alpha }{\left( 1+\alpha \beta \right) }\text{ }\right) 
\overrightarrow{\varphi }\left( x\right) =0.  \label{M30}
\end{equation}%
By analogy we obtain the equation for $\overrightarrow{\psi }$

\begin{equation}
\left( D-\frac{\alpha }{\left( 1-\alpha \beta \right) }\right) 
\overrightarrow{\psi }\left( x\right) =0.  \label{M43}
\end{equation}%
Introducing the notations 
\begin{equation*}
\alpha _{1}=\frac{\alpha }{(1+\alpha \beta )},\qquad \alpha _{2}=\frac{%
\alpha }{(1-\alpha \beta )}
\end{equation*}%
we rewrite the equations (\ref{M30}) and (\ref{M43}) in the form

\begin{equation}
\left( D+\alpha _{1}\right) \overrightarrow{\varphi }\left( x\right) =0
\label{M63}
\end{equation}%
and 
\begin{equation}
\left( D-\alpha _{2}\right) \overrightarrow{\psi }\left( x\right) =0.
\label{M64}
\end{equation}
When $\beta =0$ we arrive at the quaternionic form of the Maxwell equations
in the nonchiral case, but in general the wave numbers $\alpha _{1}$ and $%
\alpha _{2}$ are different and physically characterize the propagation of
waves of opposing circular polarizations.

As was shown in preceding sections, the functions $\overrightarrow{\varphi }$
and $\overrightarrow{\psi }$ can be approximated by right linear
combinations of functions $\left\{ \mathcal{K}_{\alpha _{1},n}^{\pm
}\right\} _{n=1}^{\infty }$ and $\left\{ \mathcal{K}_{-\alpha _{2},n}^{\pm
}\right\} _{n=1}^{\infty }$ respectively. The vectors $\overrightarrow{E}$
and $\overrightarrow{H}$ are easily obtained from $\overrightarrow{\varphi }$
and $\overrightarrow{\psi }$:%
\begin{equation}
\overrightarrow{E}=\frac{1}{2}(\overrightarrow{\varphi }+\overrightarrow{%
\psi })\qquad \text{\ and \qquad }\overrightarrow{H}=\frac{1}{2i}(%
\overrightarrow{\varphi }-\overrightarrow{\psi }).  \label{relEH}
\end{equation}%
Consequently, all the results of preceding sections are applicable to the
electromagnetic field. As before we start with exterior domains. The
radiation condition for the vectors $\overrightarrow{E}$ and $%
\overrightarrow{H}$ is the Silver-M\"{u}ller condition:%
\begin{equation}
\overrightarrow{E}-\left[ \frac{x}{\left| x\right| }\times \overrightarrow{H}%
\right] =o(\frac{1}{\left| x\right| }),  \label{SM1}
\end{equation}%
or in an equivalent form%
\begin{equation}
\overrightarrow{H}+\left[ \frac{x}{\left| x\right| }\times \overrightarrow{E}%
\right] =o(\frac{1}{\left| x\right| }),  \label{SM2}
\end{equation}%
uniformly for all directions.

Note that (\ref{SM1}) and (\ref{SM2}) are fulfilled automatically if as
before $\overrightarrow{\varphi }$ satisfies (\ref{rad+}) and $%
\overrightarrow{\psi }$ satisfies (\ref{rad-}). We have 
\begin{equation*}
\overrightarrow{E}=\frac{1}{2}(\overrightarrow{\varphi }+\overrightarrow{%
\psi })=\frac{1}{2}(-\frac{ix}{\left| x\right| }\cdot \overrightarrow{%
\varphi }+\frac{ix}{\left| x\right| }\cdot \overrightarrow{\psi })+o(\frac{1%
}{\left| x\right| })
\end{equation*}%
\begin{equation*}
=\frac{x}{\left| x\right| }\cdot \frac{1}{2i}(\overrightarrow{\varphi }-%
\overrightarrow{\psi })+o(\frac{1}{\left| x\right| })=\frac{x}{\left|
x\right| }\cdot \overrightarrow{H}+o(\frac{1}{\left| x\right| }).
\end{equation*}%
The vector part of this equality gives us (\ref{SM1}) and the scalar is a
simple consequence of (\ref{SM2}). Starting with $\overrightarrow{H}$
instead of $\overrightarrow{E}$ we arrive at (\ref{SM2}).

From Theorem \ref{PSf}, equalities (\ref{relEH}) and the last observation
concerning the relation between the radiation conditions for $%
\overrightarrow{\varphi }$ and $\overrightarrow{\psi }$ from one side and
for $\overrightarrow{E}$ and $\overrightarrow{H}$ from the other, we obtain
the following criterion.

\begin{theorem}
Let complex vectors $\overrightarrow{e}$ and $\overrightarrow{h}$ belong to $%
H^{s}(\Gamma )$, $s>0$. Then in order for $\overrightarrow{e}$ and $%
\overrightarrow{h}$ to be boundary values of $\overrightarrow{E}$ and $%
\overrightarrow{H}$ satisfying the Maxwell equations (\ref{M12}) and (\ref%
{M13}) in $\Omega ^{-}$ and (\ref{SM1}) at infinity, the following condition
is necessary and sufficient%
\begin{equation}
(\overrightarrow{e}+i\overrightarrow{h})\in \func{im}Q_{\alpha _{1}}\text{%
\qquad and\qquad }(\overrightarrow{e}-i\overrightarrow{h})\in \func{im}%
Q_{-\alpha _{2}}.  \label{criter}
\end{equation}
\end{theorem}

Now from Theorem \ref{complext} we obtain immediately the following
important result opening the possibility to apply the systems of
quaternionic fundamental solutions $\left\{ \mathcal{K}_{\alpha
_{1},n}^{-}\right\} _{n=1}^{\infty }$ and $\left\{ \mathcal{K}_{-\alpha
_{2},n}^{-}\right\} _{n=1}^{\infty }$ to approximation of the
electromagnetic field in exterior domains.

\begin{theorem}
Let both $\alpha _{1}^{2}$ and $\alpha _{2}^{2}$ be not eigenvalues of the
Dirichlet problem in $V$. Then if (\ref{criter}) is fulfilled the vectors $%
\overrightarrow{e}$ and $\overrightarrow{h}$ belonging to $H^{s}(\Gamma )$,\ 
$s>1$ can be approximated with an arbitrary precision (in the norm of $%
H^{s-1}(\Gamma )$) by right linear combinations of the form%
\begin{equation*}
\overrightarrow{e}_{N}=\frac{1}{2}(\sum_{j=1}^{N}\mathcal{K}_{\alpha
_{1},j}^{-}a_{j}+\sum_{j=1}^{N}\mathcal{K}_{-\alpha _{2},j}^{-}b_{j})
\end{equation*}%
and%
\begin{equation*}
\overrightarrow{h}_{N}=\frac{1}{2i}(\sum_{j=1}^{N}\mathcal{K}_{\alpha
_{1},j}^{-}a_{j}-\sum_{j=1}^{N}\mathcal{K}_{-\alpha _{2},j}^{-}b_{j}),
\end{equation*}%
where $a_{j}$ and $b_{j}$ are constant complex quaternions.
\end{theorem}

\begin{proof}
Due to Theorem \ref{complext} we have that there exist such $a_{j}$ and $%
b_{j}$ that 
\begin{equation*}
\left\| \overrightarrow{e}+i\overrightarrow{h}-\sum_{j=1}^{N}\mathcal{K}%
_{\alpha _{1},j}^{-}a_{j}\right\| _{H^{s-1}(\Gamma )}<\varepsilon
\end{equation*}%
and 
\begin{equation*}
\left\| \overrightarrow{e}-i\overrightarrow{h}-\sum_{j=1}^{N}\mathcal{K}%
_{-\alpha _{2},j}^{-}b_{j}\right\| _{H^{s-1}(\Gamma )}<\varepsilon .
\end{equation*}%
From these two inequalities we obtain the necessary result.
\end{proof}

In a similar way we obtain the corresponding result for interior domains.

\begin{theorem}
Let both $\alpha _{1}^{2}$ and $\alpha _{2}^{2}$ be not eigenvalues of the
Dirichlet problem in $\Omega ^{+}$ and the following condition (which is a
necessary and sufficient condition of the extendability of the vectors $%
\overrightarrow{e}$ and $\overrightarrow{h}$ into $\Omega ^{+}$ in such a
way that their extensions satisfy (\ref{M12}) and (\ref{M13})) be fulfilled%
\begin{equation*}
(\overrightarrow{e}+i\overrightarrow{h})\in \func{im}P_{\alpha
_{1}}(H^{s}(\Gamma ))\text{\qquad and\qquad }(\overrightarrow{e}-i%
\overrightarrow{h})\in \func{im}P_{-\alpha _{2}}(H^{s}(\Gamma )),\qquad s>1.
\end{equation*}%
Then $\overrightarrow{e}$ and $\overrightarrow{h}$ can be approximated with
an arbitrary precision (in the norm of $H^{s-1}(\Gamma )$) by right linear
combinations of the form%
\begin{equation*}
\overrightarrow{e}_{N}=\frac{1}{2}(\sum_{j=1}^{N}\mathcal{K}_{\alpha
_{1},j}^{+}a_{j}+\sum_{j=1}^{N}\mathcal{K}_{-\alpha _{2},j}^{+}b_{j})
\end{equation*}%
and%
\begin{equation*}
\overrightarrow{h}_{N}=\frac{1}{2i}(\sum_{j=1}^{N}\mathcal{K}_{\alpha
_{1},j}^{+}a_{j}-\sum_{j=1}^{N}\mathcal{K}_{-\alpha _{2},j}^{+}b_{j}),
\end{equation*}%
where $a_{j}$ and $b_{j}$ are constant complex quaternions.
\end{theorem}

\section{Numerical realization}

Let $\Gamma =\partial \Omega ^{-}$ be a closed sufficiently smooth surface
in $\mathbb{R}^{3}.$ Consider the following exterior boundary value problem
for the Maxwell equations%
\begin{equation}
\limfunc{rot}\overrightarrow{E}(x)=-i\alpha (\overrightarrow{H}(x)+\beta 
\limfunc{rot}\overrightarrow{H}(x)),\qquad x\in \Omega ^{-},  \label{prob1}
\end{equation}%
\begin{equation}
\limfunc{rot}\overrightarrow{H}(x)=i\alpha (\overrightarrow{E}(x)+\beta 
\limfunc{rot}\overrightarrow{E}(x)),\qquad x\in \Omega ^{-},  \label{prob2}
\end{equation}%
\begin{equation}
\left[ \overrightarrow{E}(x)\times \overrightarrow{n}(x)\right] =%
\overrightarrow{f}(x),\qquad x\in \Gamma ,  \label{prob3}
\end{equation}%
where $\overrightarrow{f}(x)$ is a given tangential field and at infinity
the vectors $\overrightarrow{E}$ and $\overrightarrow{H}$ satisfy the\
Silver-M\"{u}ller radiation condition (\ref{SM1}) or (\ref{SM2}). As before $%
\overrightarrow{n}$ \ stands for the unit outward normal to $\Gamma .$

Using results of the preceding section this problem can be rewritten in
following equivalent form%
\begin{equation}
\left( D+\alpha _{1}\right) \varphi (x)=0,\qquad x\in \Omega ^{-},
\label{d1}
\end{equation}%
\begin{equation}
\left( D-\alpha _{2}\right) \psi (x)=0,\qquad x\in \Omega ^{-},  \label{d2}
\end{equation}%
\begin{equation}
\frac{1}{2}\left[ \left( \varphi (x)+\psi (x)\right) \times \overrightarrow{n%
}(x)\right] =\overrightarrow{f}(x),\qquad x\in \Gamma ,  \label{d3}
\end{equation}%
\begin{equation}
\limfunc{Sc}\varphi (x)=0,\qquad \limfunc{Sc}\psi (x)=0,\qquad x\in \Gamma ,
\label{Scalbound}
\end{equation}

and at infinity the functions $\varphi (x)$ and $\psi (x)$ satisfy the
conditions%
\begin{equation}
\left( 1+\frac{ix}{\left| x\right| }\right) \cdot \varphi (x)=o\left( \frac{1%
}{\left| x\right| }\right) ,\qquad \left| x\right| \rightarrow \infty ,
\label{d4}
\end{equation}%
\begin{equation}
\left( 1-\frac{ix}{\left| x\right| }\right) \cdot \psi (x)=o\left( \frac{1}{%
\left| x\right| }\right) ,\qquad \left| x\right| \rightarrow \infty .
\label{d5}
\end{equation}%
Note that due to the uniqueness of the solution of the exterior Dirichlet
problem for the Helmholtz equation condition (\ref{Scalbound}) implies that $%
\varphi $ and $\psi $ will be purely vectorial on the whole domain $\Omega
^{-}$.

We look for an approximate solution of (\ref{d1})-(\ref{d5}) in the form%
\begin{equation}
\varphi _{N}(x)=\sum_{j=1}^{N}\mathcal{K}_{\alpha _{1},j}^{-}(x)a_{j}\qquad 
\text{and\qquad }\psi _{N}(x)=\sum_{j=1}^{N}\mathcal{K}_{-\alpha
_{2},j}^{-}(x)b_{j},  \label{approxsol}
\end{equation}%
where $a_{j}$ and $b_{j}$ are constant complex quaternions.

In order to find the coefficients $a_{j}\ $and $b_{j}$ we use the
collocation method. In (\ref{approxsol}) we have $8N$ unknown complex
quantities, the components of $a_{j}$ and $b_{j}$. \ Thus it is necessary to
obtain $8N$ linearly independent equations with respect to components of $%
a_{j}$ and $b_{j}$.

Every collocation point generates four equations, two of which correspond to
the boundary condition (\ref{d3}) and the other two correspond to (\ref%
{Scalbound}). Consequently in order to determine the coefficients $a_{j}$
and $b_{j}$ we need $2N$ collocation points. After having solved the
corresponding system of linear algebraic equations we obtain the approximate
solution of the problem (\ref{prob1})-(\ref{prob3}):

\begin{equation*}
\overrightarrow{E}(x)=\frac{1}{2}\left( \varphi _{N}(x)+\psi _{N}(x)\right)
,\qquad x\in \Omega ^{-},
\end{equation*}%
\begin{equation*}
\overrightarrow{H}(x)=\frac{1}{2i}\left( \varphi _{N}(x)-\psi _{N}(x)\right)
,\qquad x\in \Omega ^{-}.
\end{equation*}%
\ The method described above was tested using the following exact solution.

Let $\beta =0$ and consequently $\alpha =\alpha _{1}=\alpha _{2}$. The
vectors

\begin{equation*}
\overrightarrow{E}^{m}(x)=\limfunc{rot}\overrightarrow{c}\theta _{\alpha
}(x)=\left( 
\begin{array}{c}
c_{3}\partial _{2}\theta _{\alpha }(x)-c_{2}\partial _{3}\theta _{\alpha }(x)
\\ 
c_{1}\partial _{3}\theta _{\alpha }(x)-c_{3}\partial _{1}\theta _{\alpha }(x)
\\ 
c_{2}\partial _{1}\theta _{\alpha }(x)-c_{1}\partial _{2}\theta _{\alpha }(x)%
\end{array}%
\right)
\end{equation*}%
and%
\begin{equation*}
\overrightarrow{H}^{m}(x)=-\frac{1}{i\alpha }\limfunc{rot}\overrightarrow{E}%
^{m}(x),\qquad x\in \mathbb{R}^{3}\setminus \left\{ 0\right\} ,
\end{equation*}%
where $\overrightarrow{c}\in \mathbb{R}^{3}$ is constant, represent the
electromagnetic field of a magnetic dipole situated at the origin %
\cite[Sect. 4.2]{CK1}. They satisfy (\ref{prob1}) and (\ref{prob2}) (for $%
\beta =0)$ as well as the Silver-M\"{u}ller conditions at infinity.

Let $\Gamma $ be a unit sphere with its centre at the origin. Then $%
\overrightarrow{E}^{m}$ and $\overrightarrow{H}^{m}$ give us the solution of
the following boundary value problem%
\begin{equation*}
\limfunc{rot}\overrightarrow{E}(x)=-i\alpha \overrightarrow{H}(x),\qquad
x\in \Omega ^{-},
\end{equation*}%
\begin{equation*}
\limfunc{rot}\overrightarrow{H}(x)=i\alpha \overrightarrow{E}(x),\qquad x\in
\Omega ^{-},
\end{equation*}

\begin{equation*}
\left[ \overrightarrow{E}(x)\times \overrightarrow{n}(x)\right] =%
\overrightarrow{f}(x),\qquad x\in \Gamma
\end{equation*}%
where 
\begin{equation*}
\overrightarrow{f}(x)=\left[ \left( 
\begin{array}{c}
c_{3}\partial _{2}\theta _{\alpha }(x)-c_{2}\partial _{3}\theta _{\alpha }(x)
\\ 
c_{1}\partial _{3}\theta _{\alpha }(x)-c_{3}\partial _{1}\theta _{\alpha }(x)
\\ 
c_{2}\partial _{1}\theta _{\alpha }(x)-c_{1}\partial _{2}\theta _{\alpha }(x)%
\end{array}%
\right) \times \overrightarrow{n}(x)\right] .
\end{equation*}

As the auxiliary surface $\Gamma ^{-}$ containing points $y_{n}^{-}$ we have
chosen the sphere with centre at the origin and radius 0.15. In the
following table we present the results for different values of $N$. The
corresponding errors represent the absolute maximum difference between the
exact and the approximate solutions at the points on the sphere with centre
at the origin and radius 5.

\begin{center}
\begin{tabular}{||c|c|c||}
\hline\hline
$N$ & Error for $\overrightarrow{E}$ & Error for $\overrightarrow{H}$ \\ 
\hline
3 & 0.802E-02 \  & 0.756E-02 \  \\ \hline
5 & 0.346E-02 & 0.250E-02 \\ \hline
10 & 0.334E-03 & 0.299E-03 \\ \hline
15 & 0.137E-03 & 0.132E-03 \\ \hline
20 & 0.128E-04 & 0.166E-04 \\ \hline
25 & 0.187E-04 & 0.149E-04 \\ \hline
30 & 0.465E-05 & 0.588E-05 \\ \hline
35 & 0.278E-06 & 0.398E-06 \\ \hline\hline
\end{tabular}
\end{center}

A quite fast convergence of the method can be appreciated (all numerical
results were obtained on a PC Pentium 3).

Let us notice that the approximation by linear combinations of quaternionic
fundamental solutions can be applied to other classes of boundary value
problems for the Maxwell system like for example the impedance problem.

\begin{acknowledgement}
This work was supported by CONACYT Project 32424-E, Mexico. The authors
express their gratitude to Prof. Nikolai Tarkhanov for helpful discussions.
\end{acknowledgement}

\end{document}